\documentclass[aps,pra,amssymb,a4paper]{revtex4}
\usepackage{graphicx}

\normalsize
\begin{document}
\title{\bf Measurable entanglement}

\author{ Alexander Klyachko, Bar{\i}\c{s} \"{O}ztop,
Alexander S. Shumovsky}

\affiliation{Faculty of Science, Bilkent University, Bilkent,
Ankara, 06800, Turkey}

\begin{abstract}
Amount of entanglement carried by a quantum bipartite state is
usually evaluated in terms of concurrence (see Ref. 1). We give a
physical interpretation of concurrence that reveals a way of its
direct measurement and discuss possible generalizations.
\end{abstract}


\maketitle

Entanglement, which has been considered for decades in the context
of fundamentals of quantum mechanics, turns now more and more into
a key tool of practical realization of quantum information
technologies. The quantum key distribution$^2$ for completely
secured communications should be mentioned here first of all
(e.g., see Ref. 3).

The design and manufacturing of generators of entangled states
require a control of amount of entanglement carried by the states.
For pure state $\psi$ of bipartite systems
$\mathcal{H}_A\otimes\mathcal{H}_B$ of format $n\times n$ ($n=
\dim \mathcal{H}_{A,B}$; $n=2$ corresponds to qubits, $n=3$
corresponds to qutrits, etc.), this quantity is given by the {\it
concurrence}
\begin{eqnarray}
C(\psi)= \sqrt{ \nu [1-\mbox{Tr}( \rho_r^2)]}, \label{conc}
\end{eqnarray}
that has been proposed in Ref. 1.
Here $\rho_r$ denotes the reduced (single-party) density matrix,
corresponding to the state $\psi$, and we use the normalization
factor $\nu=\frac{n}{n-1}$, to reduce the concurrence to the
interval $[0,1]$. See Ref. 4 for further discussion.

The aim of this note is to give a natural physical interpretation
of the concurrence and thus to show a way of direct measurement of
the amount of bipartite entanglement in terms of mean values of
certain physical quantities. Our approach also suggests a general
definition of the concurrence for multipartite systems discussed
below.

It has been shown in Ref. 7 that entanglement like coherence and
squeezing can be associated with
quantum fluctuations or quantum uncertainties, which are minimal
for coherent (separable) states and maximal for completely
entangled states. The fluctuations are measured by the {\it total
variance\,} defined by equation
\begin{eqnarray}
V(\psi) = \sum_\alpha \langle \psi|X_\alpha^2|\psi \rangle -
\langle
\psi|X_\alpha|\psi \rangle^2, \label{2}
\end{eqnarray}
where the sum is extended over orthonormal basis $X_\alpha$ of Lie
algebra of local observables. The {\it basic observables\,}
$X_\alpha$ act on one of the components: $X_\alpha=X^A_i$ or
$X_\alpha=X^B_j$, where $X^A_i$ and $X^B_j$ are are orthonormal
bases in the space of traceless Hermitian operators in
$\mathcal{H}_A$ and $\mathcal{H}_B$ respectively. It is important
to realize that the total variance is independent of the choice of
the basic observables $X_\alpha$.

The total uncertainty of all basic observables attains its maximal
value in the case of completely entangled states (like Bell states
of two qubits).



The first sum in the right-hand side of Eq. (2) is independent of
the state $\psi$. In fact, the sum
\begin{eqnarray}
C=\sum_\alpha X_\alpha^2 \nonumber
\end{eqnarray}
known as {\it Casimir operator\,}$^8$, acts as a multiplication by
a scalar $C_{AB}$ (equal to $6$ for two qubits, for example).
Thus,
\begin{eqnarray}
V(\psi)=C_{AB}- \sum_i \langle \psi |X_i| \psi \rangle^2,
\label{3}
\end{eqnarray}
so that the measurement of the total uncertainty is reduced to the
measurement of mean values of basic observables in the right-hand
side of Eq. (3). In the case of complete entanglement
\begin{eqnarray}
\langle \psi |X_\alpha| \psi \rangle =0 \nonumber
\end{eqnarray}
for all $\alpha$ (see Ref. 7), so that the total uncertainty (3)
achieves its maximum.

We now show that concurrence (1) can be equivalently expressed in
terms of the total uncertainty (3) in the case of bipartite
systems. Consider first the case of two qubits with the state
\begin{eqnarray}
|\psi \rangle = \sum_{\ell , \ell' =0}^1 \psi_{\ell \ell'} |\ell ,
\ell ' \rangle , \quad \sum_{\ell , \ell'=0}^1 |\psi_{\ell
\ell'}|^2=1, \label{4}
\end{eqnarray}
where $|\ell , \ell' \rangle \equiv |\ell \rangle \otimes |\ell'
\rangle$ denotes a composite state. It can be easily seen that the
concurrence (1) is then cast to the form
\begin{eqnarray}
C(\psi) & = & 2|\psi_{00} \psi_{11} - \psi_{01} \psi_{10}| \nonumber \\
& = & 2[|\psi_{00}|^2|\psi_{11}|^2+|\psi_{01}|^2|\psi_{10}|^2
\nonumber
\\ & - & 2Re(\psi_{00}\psi_{11}\psi_{01}^*\psi_{10}^*)]^{1/2} \label{5}
\end{eqnarray}
On the other hand using Pauli operators
\begin{eqnarray}
\sigma_x & = & |0 \rangle \langle 1|+|1 \rangle \langle 0|,
\nonumber \\ \sigma_y & = & -i(|0 \rangle \langle 1|-|1 \rangle
\langle 0|), \label{6} \\ \sigma_z & = & |0 \rangle \langle 0|-|1
\rangle \langle 1| \nonumber
\end{eqnarray}
as the basic local observables $X^A_i$ and $X^B_j$ one gets
\begin{eqnarray}
V(\psi) & = &
4+4[|\psi_{00}|^2|\psi_{11}|^2+|\psi_{01}|^2|\psi_{10}|^2
\nonumber \\ & - & 2Re(\psi_{00}\psi_{11}\psi_{01}^*\psi_{10}^*)].
\label{7}
\end{eqnarray}
Comparing now Eqs. (5) and (7) and taking into account that
$V_{max}=6$ and $V_{min}=4$ in the case of completely entangled
and unentangled states of two qubits, respectively, we get
\begin{eqnarray}
C(\psi) = \sqrt{\frac{V(\psi)-V_{\min}}{V_{\max}-V_{\min}}}
\label{var_conc}
\end{eqnarray}
in the case of the general two-qubit state (4). Thus, the amount
of entanglement carried by a pure two-qubit state can be
determined by measurement of mean values of the basic observables
given by Pauli operators (6). These observables can be directly
measured in experiments, say by the Stern-Gerlach apparatus in the
case of spins, or by means of polarizers in the case of photons,
etc.$^9$.

As a matter of fact, this expression (8) is equivalent to (1) for
any bipartite system$^{10}$.
For example, in the representation of
basic observables for qutrits $(n=3)$ given in Ref. 5, the maximal
and minimal values of total uncertainty in bipartite system are
$V_{max}=32/3$ and $V_{min}=8$, respectively. A possible
realization of  qutrits is provided by biphotons$^{11}$.

Eq. (8) allows us to interpret concurrence (1) as a square root of
the normalized total uncertainty of basic observables, specifying
the system. In view of Eq. (3), the latter can be determined in
terms of measurement of expectation values of the basic
observables $\langle \psi|X_\alpha|\psi \rangle$. In other words,
Eq. (8) provides an {\it operational} definition of measure of
bipartite entanglement. Note also that equation (8) allows to
define the concurrence for any multipartite system.

Our consideration so far have applied to the pure bipartite
states. In connection with mixed states, we now note that the
uncertainty of an observable $X_i$ can be interpreted as a
specific Wigner-Yanase ``quantum information" about a state $\psi$
extracted from the macroscopic measurement of $X_i$ in this
state$^{12}$. The generalization of Wigner-Yanase ``information"
on the case of mixed states with the density matrix $\rho$ has the
form
\begin{eqnarray}
I_i(\rho)=- \frac{1}{2} Tr([X_i, \rho^{1/2}]^2) \geq 0 . \label{9}
\end{eqnarray}
It can be easily seen that in the case of pure states when
$\rho=|\psi \rangle \langle \psi|$ the total amount of
Wigner-Yanase skew information
\begin{eqnarray}
I(\rho)= \sum_i I_i(\rho) \label{10}
\end{eqnarray}
coincides with the total uncertainty (2). The supposition is that
Eq. (10) can represent a reasonable estimation from above for the
amount of concurrence in the mixed bipartite state$^{10}$.


\begin{acknowledgments}
Bar{\i}\c{s} \"{O}ztop would like to acknowledge the Scientific
and Technical Research Council of Turkey (T\"UB\.ITAK) for
financial support.
\end{acknowledgments}




\begin{references}
\bibitem{1} S. Hill and W.K. Wootters, Phys. Rev. Lett.
{\bf 78}, 5022 (1997); P. Rungta, V. Bu\v{z}ek, C.M. Caves, M.
Hillery, and  G. J. Milburn, Phys. Rev. A {\bf64}, 042315 (2001).
\bibitem{2} C.H. Bennett and G. Brassard, in {\it Proceedings of
the IEEE Int. Conf. on Computers, Systems, and Signal Processing}
(IEEE, New York, 1984), p. 175; A. Ekert, Phys. Rev. Lett. {\bf
67}, 1661 (1991).
\bibitem{3} J. Ouellette, The Industrial Physicist, {\bf 10}, 22
(2004)
\bibitem{4} F. Minnert, M. Ku\'s and A.~Buchleitner,
Phys. Rev. Lett. 92 (2004), 167902.
\bibitem{5} C.M. Caves and G.J. Milburn,
Optics Commun. {\bf 179}, 439 (2000).
\bibitem{6} D. Kaszlikowski, D.K.L. Oi, M. Christandl, K. Chang, A.
Ekert, L.C. Kwek, and C.H. Oh, Phys. Rev. A {\bf 67}, 012310
(2003); T. Durt T., N. Cerf, N. Gisin, and M. Zukowski, Phys. Rev.
A, 67, 012311, (2003); T. Durt, D. Kaszlikowski, J.-L. Chen, and
L.C. Kwek, Phys. Rev. A {\bf 69}, 032313 (2004).

\bibitem{7} M.A. Can, A.A. Klyachko, and A.S. Shumovsky, Phys.
Rev. A {\bf 66}, 02111 (2002); A.A. Klyachko and A.S. Shumovsky,
J. Opt. B: Quant. and Semiclass. Opt. {\bf 5}, S322 (2003); A.A.
Klyachko and A.S. Shumovsky, J. Opt. B: Quant. and Semiclass. Opt.
{\bf 6}, S29 (2004).
\bibitem{8} A. Bohm, {\it
Quantum mechanics: foundations and applications} (
Springer-Verlag, New York,    1993).
\bibitem{9} M. Nielsen and I. Chuang, {\it Quantum
Computation and Quantum Information} (Cambridge University Press,
New York, 2000).
\bibitem{10} A.A. Klyachko, B. \"{O}ztop, and A.S. Shumovsky, {\it
to be published}.
\bibitem{11} A.V. Burlakov, M.V. Chechova, O.A. Karabutova, D.N.
Klyshko, and S.P. Kulik, Phys. Rev. A {\bf 60}, R4209 (1999); Yu.
I. Bogdanov, M.V. Chekhova, S.P. Kulik, G.A. Maslennikov, A.A.
Zhukov, C.H. Oh, and M.K. Tey, Phys. Rev. Lett. {\bf 93}, 230503
(2004).
\bibitem{12} E.P Wigner, Z. Physik {\bf 131}, 101 (1952); E.P. Wigner and M.M. Yanase,
Proc. Nat. Acad. Sci. USA {\bf 19},
910 (1963).

\end{references}
\end{document}